\documentclass{appolb}
\usepackage{graphicx}

\usepackage[english]{babel}
\usepackage[utf8]{inputenc}
\usepackage[T1]{fontenc}

\usepackage{amsfonts}
\usepackage{amssymb}

\usepackage{tikz}
\usepackage{nicefrac}
\usepackage{graphicx}
\usetikzlibrary{decorations.pathmorphing} 
\usetikzlibrary{decorations.pathreplacing}
\usetikzlibrary{decorations.markings}
\usetikzlibrary{arrows} 
\usepackage{multirow}
\usepackage{bbold}
\usepackage{cite}

\definecolor{DRed}{RGB}{139,0,0}
\definecolor{DBlue}{RGB}{0,0,139}

\newcommand{\dd}{\,\mathrm{d}}
\newcommand{\ee}{\,\mathrm{e}}

\begin{document} 

\title{Finite Temperature Lattice QCD - Baryons in the Quark-Gluon Plasma}  
\author{Gert Aarts, Chris Allton, Davide De Boni, Simon Hands, Benjamin
Jäger\thanks{Presented at the Workshop “Excited QCD 2016”, Portugal, 6-12 March
2016.}, 
Chrisanthi Praki, 
\address{Department of Physics, College of Science,
Swansea University,\newline Swansea SA2 8PP, United Kingdom}
\newline\phantom{stuff}\newline 
{Jon-Ivar Skullerud}
\address{Department of Mathematical Physics, National University of Ireland
Maynooth, Maynooth, County Kildare, Ireland\\
School of Mathematics, Trinity College, Dublin 2, Ireland} }
\maketitle 

\begin{abstract}
Baryonic correlation functions provide an ideal tool to study parity doubling
and chiral symmetry using lattice simulations. We present a study using $2+1$
flavors of anisotropic Wilson clover fermions on the FASTSUM ensembles and
find clear evidence that parity doubling emerges in the quark-gluon plasma.
This result is confirmed on the level of spectral functions, which are obtained using a
MEM reconstruction. We further highlight the importance of Gaussian smearing in this study.
\end{abstract}
  
\section{Introduction}

Symmetries are essential to understand interactions in nature and have led to
many discoveries in the past. Here we study baryons at nonzero temperature,
for which, in contrast to mesons, not many lattice studies are
available~\cite{DeTar:1987ar,DeTar:1987xb,Datta:2012fz,Pushkina:2004wa}.
We focus in particular on parity doubling and chiral symmetry restoration, 
which are expected to coincide for a phase where chiral symmetry is manifest. In a previous
study~\cite{Aarts:2015mma,Aarts:2015xua}, we focused on correlation function
itself; here we extend our study to include spectral functions and different levels of smearing.

\section{Setup}

We use non-perturbatively $\mathcal{O}(a)$ improved, anisotropic Wilson
fermions with $2+1$ flavors on configurations generated by the FASTSUM
collaboration~\cite{Amato:2013naa,Aarts:2014nba,Aarts:2014cda}, based on the
parameters of the Hadron Spectrum Collaboration~\cite{Edwards:2008ja}. The simulation parameter are listed in
table~\ref{Tab1}.
\begin{table}[!ht]
\centering
\begin{tabular}{cccccc}
	$N_s$ & $N_t$ & $T [\mathrm{MeV}]$ & $T/T_c$ & $N_{\mathrm{cfg}}$ &
	$N_{\mathrm{src}}$\\
\hline 
24 &128& 44 &0.24 &171 &2 \\
24 &40 &141 &0.76 &301 &4 \\
24 &36 &156 &0.84 &252 &4 \\
24 &32 &176 &0.95 &1000 &2 \\
24 &28 &201 &1.09 &501 &4 \\
24 &24 &235 &1.27 &1001 &2 \\
24 &20 &281 &1.52 &1000 &2 \\
24 &16 &352 &1.90 &1001 &2\\
\hline
\end{tabular}
\caption{\label{Tab1} Simulation parameters. }
\end{table}
These ensembles span a wide range in temperatures, ranging from
$44\,\mathrm{MeV}$ to $352\,\mathrm{MeV}$, and in terms of the critical 
temperature from $0.24$ to $1.9 \, T_c$. We use a fixed spatial
lattice spacing of $a_s = 0.1227(8)\,\mathrm{fm}$ and a finer lattice spacing in
the time directions, such that the anisotropy is $a_s/a_t = 3.5$~\cite{Lin:2008pr}.
The strange quark mass has been tuned to its physical value, while the light quarks remain
heavier than in nature, which results in a pion mass of
$384(4)\,\mathrm{MeV}$~\cite{Lin:2008pr}. Further details of the ensembles can
be found in~\cite{Amato:2013naa,Aarts:2014nba,Aarts:2014cda}. 

For the nucleon interpolating operator we use a standard
definition (see e.g.~\cite{Montvay:1994cy,Gattringer:2010zz}),
\begin{equation}
O_N = \epsilon_{abc} \, u_a \left( u_b \, C \gamma_5\,  d_c \right).
\end{equation}
With this definition and the projector to positive parity $P_+ =
\frac{1}{2}\left(\mathbb{1} + \gamma_4 \right)$, the correlation function of the
nucleon can be obtained by
\begin{equation}
C(t) = \sum_{\vec{x}} \,\big< O_N(\vec{x}, t) \,P_+\,
\overline{O_N}(0)\big>.
\end{equation}
To enhance the overlap with the ground state, we employ Gaussian
smearing~\cite{Gusken:1989ad} on the source and the sink operator of the
correlation functions, which will be
discussed below. It will become clear that smearing is crucial for separating
out the ground state already at early Euclidean times.
Since Wilson fermions explicitly break chiral symmetry,
we do not expect that all excited states reflect parity doubling and
hence we focus in this work on the low-energy states. The correlation 
functions have been computed using the Chroma software package~\cite{Edwards:2004sx}.
 
\section{Results}
\begin{figure}[!ht] 
\centering 
\hspace{-0.0cm} 
\begin{minipage}{0.98\linewidth}
\centering
\begin{tabular}{cc}
\includegraphics[clip, trim=0.0cm 0.0cm 0.2cm
	0.0cm,width=0.59\linewidth]{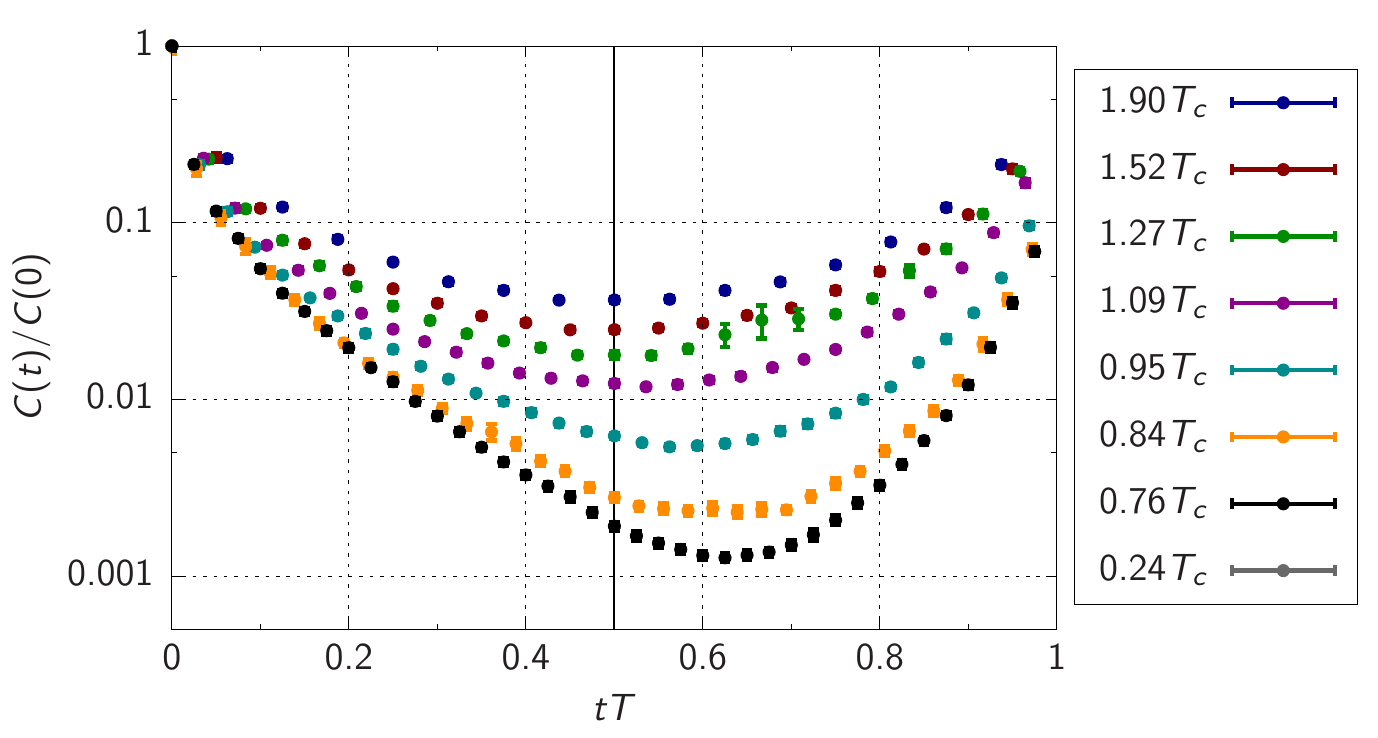} & 
\includegraphics[clip, trim=0.0cm 0.0cm 0.5cm
	0.0cm,width=0.39\linewidth]{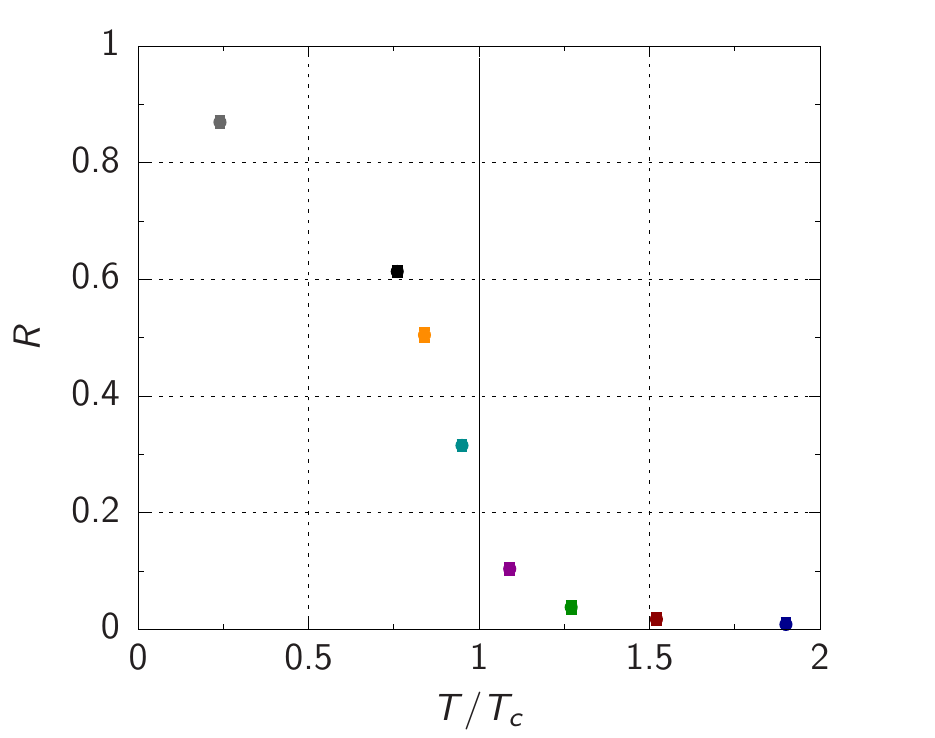}
\end{tabular}
\end{minipage}   
\caption{Left: Correlation functions of the nucleon for different
temperatures, as a function of $t\,T$. Right: The weighted average $R$, defined
in equation~\ref{Eq3}, as a function of temperature. The
error bars in both panels are of the order of the symbol size.}
\label{Fig1}
\end{figure}

The left panel of figure~\ref{Fig1} shows the correlation functions for various
temperatures. 
The individual correlation function is the result of superposition of
forward-propagating states with positive parity and backward-propagating states
with negative parity. In nature, the ground states in each channel have
different masses resulting in asymmetric correlation functions. This
behaviour is clearly reproduced in figure~\ref{Fig1} for $T < T_c$. As the
temperature the correlation function regains more and more of its reflection
symmetry around $t\,T=1/2$, which indicates
the emergence of parity doubling in the quark-gluon plasma.
To quantify this further, we look at a weighted average of ratios of correlation
functions~\cite{Datta:2012fz},
\begin{equation}
R  = \frac{1}{Z} \sum_{i=1}^{N_t/2-1} R(t_i)/\sigma_i^2, \quad
\mathrm{where}\quad  R(t) =
\frac{ C(t) - C(N_t -t)}{C(t) + C(N_t -t)} \label{Eq3}
\end{equation}
and $Z = \sum_i \sigma_i^{-2}$ is the normalization. The right panel
of figure~\ref{Fig1} shows this ratio $R$, which shows a crossover behaviour and
confirms the coincidence of parity doubling with the thermal transition.

\begin{figure}[!t] 
\centering 
\hspace{-0.0cm} 
\begin{minipage}{0.98\linewidth}
\centering
\begin{tabular}{cc}
\includegraphics[clip, trim=0.0cm 0.0cm 0.5cm
	0.0cm,width=0.49\linewidth]{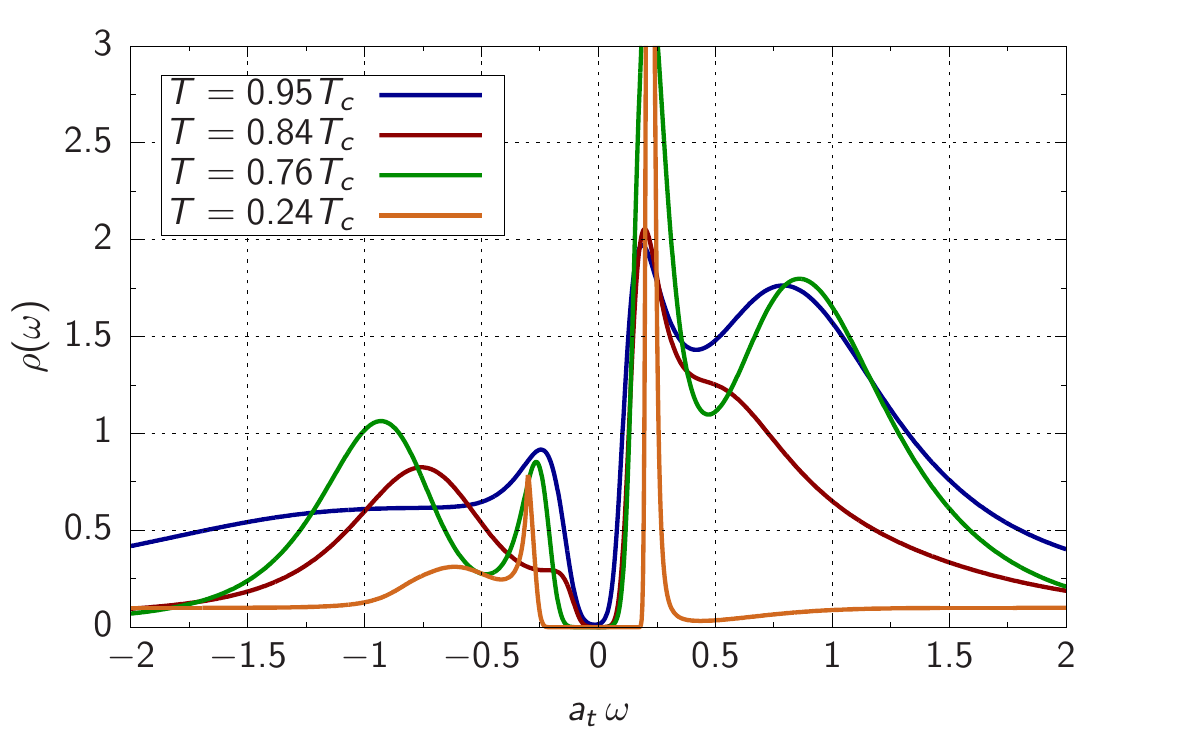} & 
\includegraphics[clip, trim=0.0cm 0.0cm 0.5cm
	0.0cm,width=0.49\linewidth]{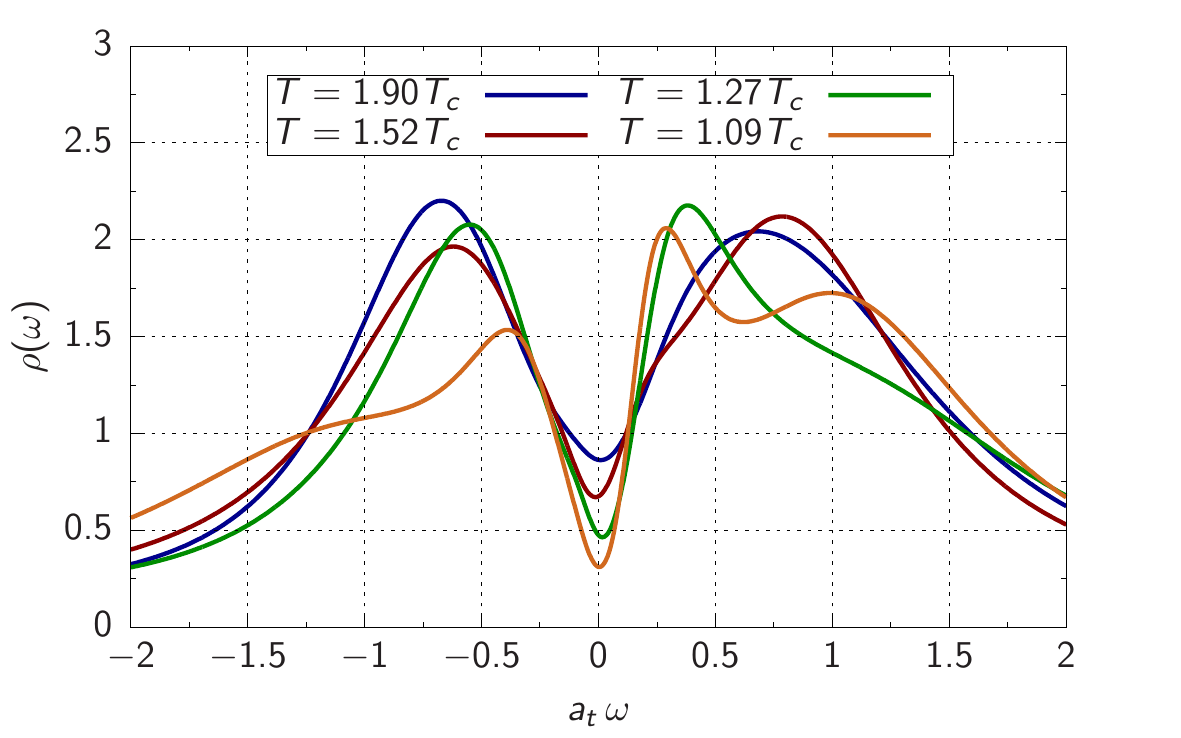}
\end{tabular} 
\end{minipage}   
\caption{The spectral function of the nucleon for a variety of temperatures 
below the critical temperature (left) and above the critical
temperature (right).}
\label{Fig2}
\end{figure}

Looking at the spectral decomposition of the correlations
functions,
\begin{equation} 
C(t) = \int_{-\infty}^{\infty} \frac{\dd \omega}{2\pi} \,K(t,\omega) \,
\rho(\omega),
\end{equation}
allows us to study properties of the nucleon system further.
The determination of the spectral function $\rho(\omega)$ is an ill-posed problem in
itself, which we study by using the Maximum Entropy
Method (MEM) ~\cite{Asakawa:2000tr} and adapting the kernel $K(t,\omega)$ to this
(fermionic) problem.
For the correlator shown in figure~\ref{Fig1}, the kernel
reads ~\cite{Praki:2015yua}
\begin{equation}
K(t,\omega) = \frac{\ee^{-\omega t}}{1+\ee^{-\omega / T}}.
\end{equation}
Figure~\ref{Fig2} shows the result of this spectral reconstruction. 
Note that the spectrum at positive (negative) $\omega$ corresponds to the
positive (negative) parity channel. At low temperatures the ground states in
both the positive and negative parity channels are clearly visible, but they
are reduced as the temperature increases. Above $T_c$ the spectral function
becomes more and more symmetric, consistent with the analysis from the
correlators directly.

\begin{figure}[!ht] 
\centering  
\hspace{-0.0cm} 
\begin{minipage}{0.98\linewidth}
\centering
\begin{tabular}{cc}
\includegraphics[clip, trim=0.0cm 0.0cm 0.5cm
	0.0cm,width=0.59\linewidth]{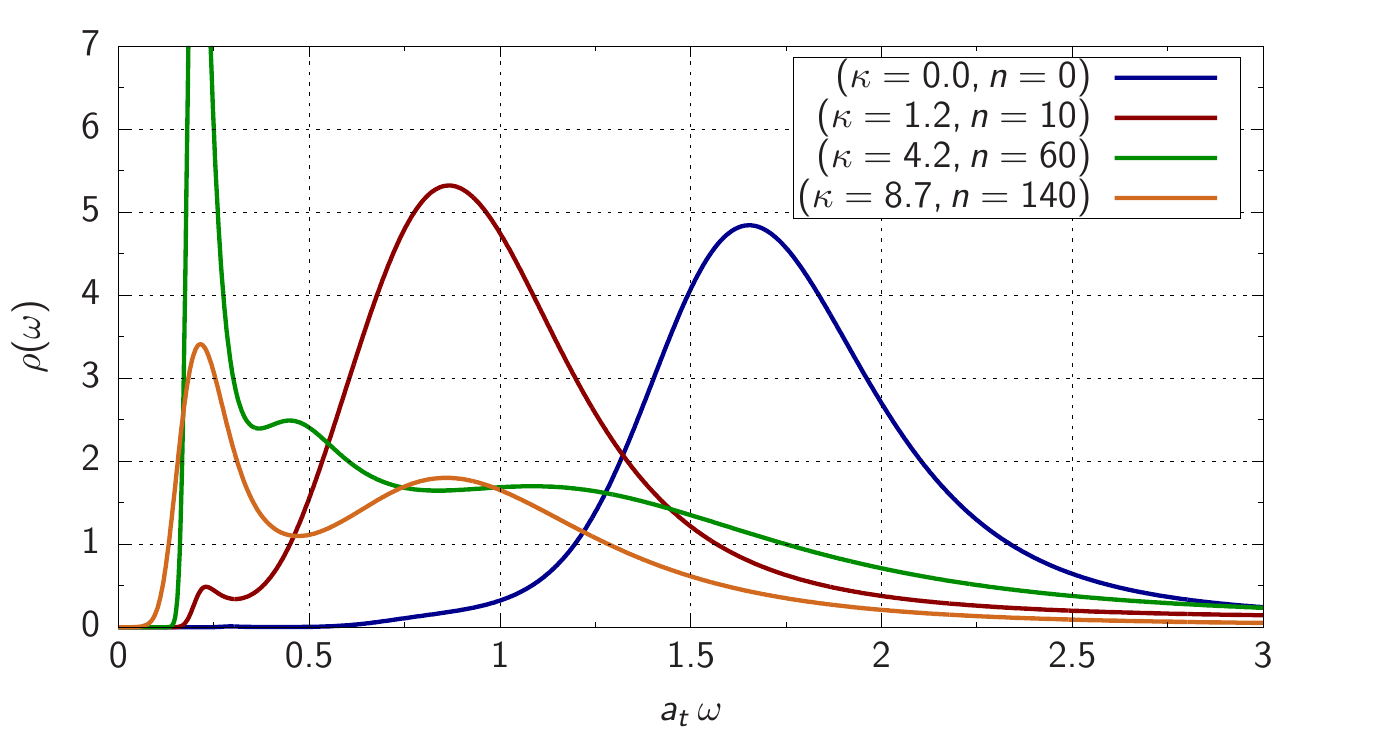} & 
\includegraphics[clip, trim=0.0cm 0.0cm 0.5cm
	0.0cm,width=0.39\linewidth]{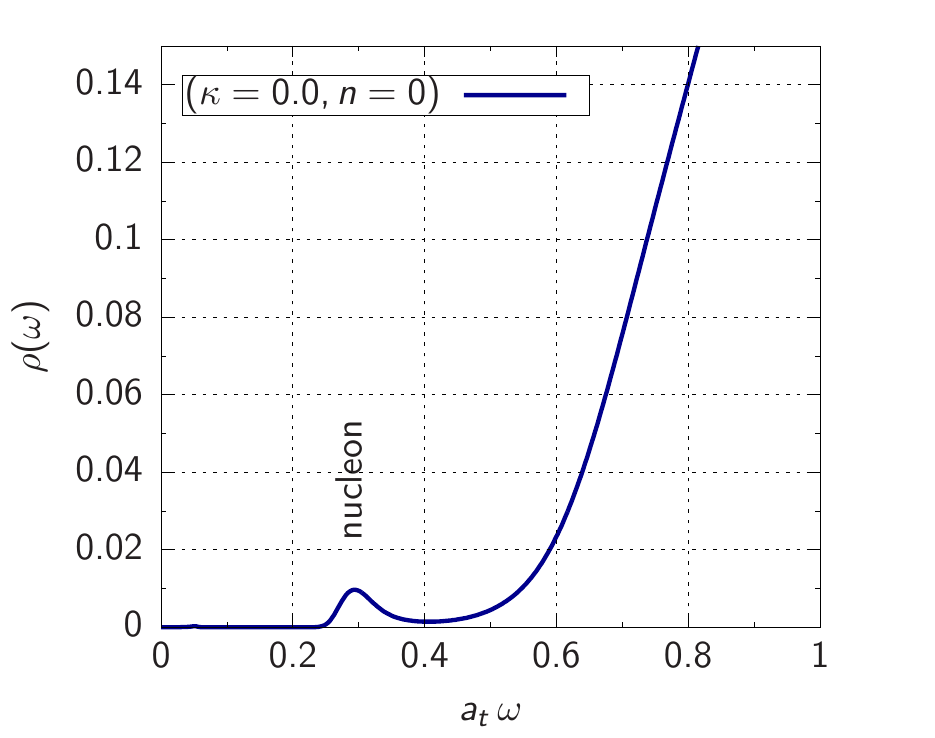}
\end{tabular} 
\end{minipage}   
\caption{The effect of smearing on one particular ensemble with $N_\tau =40$,
showing only the positive parity sector. Left: A comparison of
various levels of smearing as listed in equation~(\ref{eq8}). Right: A zoom of
the spectral function without smearing.  }
\label{Fig3}
\end{figure}

As mentioned before, we apply Gaussian smearing~\cite{Gusken:1989ad} to the
source and the sink operator, i.e.
\begin{equation}
	\psi' (x) = \frac{1}{A} \left( \mathbb{1} + \kappa \, H \right)^n \psi(x),
\end{equation}
where $H$ corresponds to the hopping part of the Dirac operator and $A$ is
an appropriate normalization. The links variables in $H$ are APE
smeared~\cite{Albanese:1987ds}. In total, we apply four different settings for
the smearing parameters $\kappa$ and $n$, 
\begin{equation}
(\kappa,n) = (0,0) ; \,(1.2,10) ; \,(4.2,60) ; \, (8.7,140), \label{eq8}
\end{equation}
to test the dependence on the smearing, which includes a setup with no smearing
at all. The data shown in the first
part of this section, i.e. figures~\ref{Fig1} and~\ref{Fig2}, have been obtained
using $(\kappa = 8.7,n=140)$. The left panel of figure~\ref{Fig3} shows
the resulting spectral functions for an ensemble with a temporal extent of $N_t=40$.
As expected the overlap with the ground state changes significantly by changing
the spectral weights with the smearing procedure. The position of the first peak,
i.e the mass of the ground state, remains however unchanged. As far as our
analysis and uncertainties allow, we also see no clear change in the width,
which still needs more data to be confirmed. In particular without any smearing
the ground state is severely suppressed, but still visible, and this is clearly
shown in the right panel of figure~\ref{Fig3}. Therefore smearing is absolutely crucial for
extracting information on the ground state. 
  
\section{Conclusion} 
We have presented a lattice study of the spin-$\frac{1}{2}$ octet nucleon in
the hadronic phase and quark-gluon plasma, spanning a wide range of temperatures
across the transition. We find clear evidence of parity doubling and thereby
chiral symmetry restoration in the quark-gluon plasma, which is further
confirmed on the level of the spectral functions. Gaussian smearing has shown
to be essential in this work. In future we will extend our study to the
spin-$\frac{3}{2}$ baryon decuplet and include different valence quark masses.

\section{Acknowledgements}
We are grateful for the computing resources made available by HPC Wales. We acknowledge
the STFC grants ST/L000369/1,    	
ST/K000411/1,  ST/H008845/1, ST/K005804/1 and ST/K005790/1,     
the STFC DiRAC HPC Facility (www.dirac.ac.uk),
PRACE grants 2011040469 and Pra05 1129q, CINECA grant INF14 FTeCP (CINECA-INFN
agreement). We thank the Royal Society, the Wolfson Foundation and the
Leverhulme Trust for their support.


\begin{thebibliography}{10}

\bibitem{DeTar:1987ar}
  C.~E.~DeTar and J.~B.~Kogut,
  Phys.\ Rev.\ Lett.\  {\bf 59} (1987) 399.
  
\bibitem{DeTar:1987xb}
 C.~E.~DeTar and J.~B.~Kogut,
  Phys.\ Rev.\ D {\bf 36} (1987) 2828.

\bibitem{Datta:2012fz}
  S.~Datta, S.~Gupta, M.~Padmanath, J.~Maiti and N.~Mathur,
  JHEP {\bf 1302} (2013) 145
  doi:10.1007/JHEP02(2013)145
  [arXiv:1212.2927 [hep-lat]].

\bibitem{Pushkina:2004wa}
  I.~Pushkina {\it et al.}  [QCD-TARO Collaboration],
  Phys.\ Lett.\ B {\bf 609} (2005) 265
  [hep-lat/0410017].


\bibitem{Aarts:2015mma}
  G.~Aarts, C.~Allton, S.~Hands, B.~Jäger, C.~Praki and J.~I.~Skullerud,
  Phys.\ Rev.\ D {\bf 92} (2015) no.1,  014503
  doi:10.1103/PhysRevD.92.014503
  [arXiv:1502.03603 [hep-lat]].

\bibitem{Aarts:2015xua}
  G.~Aarts, C.~Allton, S.~Hands, B.~Jäger, C.~Praki and J.~I.~Skullerud,
  PoS LATTICE {\bf 2015} (2015) 183
  [arXiv:1510.04040 [hep-lat]].


  
\bibitem{Amato:2013naa}
  A.~Amato, G.~Aarts, C.~Allton, P.~Giudice, S.~Hands and J.~I.~Skullerud,
  Phys.\ Rev.\ Lett.\  {\bf 111} (2013) 172001
  [arXiv:1307.6763 [hep-lat]].

\bibitem{Aarts:2014nba}
  G.~Aarts, C.~Allton, A.~Amato, P.~Giudice, S.~Hands and J.~I.~Skullerud,
  JHEP {\bf 1502} (2015) 186
  [arXiv:1412.6411 [hep-lat]].

\bibitem{Aarts:2014cda}
  G.~Aarts, C.~Allton, T.~Harris, S.~Kim, M.~P.~Lombardo, S.~M.~Ryan and J.~I.~Skullerud,
  JHEP {\bf 1407} (2014) 097
  doi:10.1007/JHEP07(2014)097
  [arXiv:1402.6210 [hep-lat]].
  
\bibitem{Edwards:2008ja}
  R.~G.~Edwards, B.~Joo and H.~W.~Lin,
  Phys.\ Rev.\ D {\bf 78} (2008) 054501
  [arXiv:0803.3960 [hep-lat]].
  
\bibitem{Lin:2008pr}
  H.~W.~Lin {\it et al.}  [Hadron Spectrum Collaboration],
  Phys.\ Rev.\ D {\bf 79} (2009) 034502
  [arXiv:0810.3588 [hep-lat]].

\bibitem{Montvay:1994cy}
  I.~Montvay and G.~M\"unster,
  ``Quantum fields on a lattice,''
  Cambridge, UK: Univ. Pr. (1994) 491 p. (Cambridge monographs on mathematical physics).

\bibitem{Gattringer:2010zz}
  C.~Gattringer and C.~B.~Lang,
  ``Quantum chromodynamics on the lattice,''
  Lect.\ Notes Phys.\  {\bf 788} (2010) 1.

\bibitem{Gusken:1989ad}
  S.~G\"usken, U.~Low, K.~H.~M\"utter, R.~Sommer, A.~Patel and K.~Schilling,
  Phys.\ Lett.\ B {\bf 227} (1989) 266.

\bibitem{Edwards:2004sx}
  R.~G.~Edwards {\it et al.}  [SciDAC and LHPC and UKQCD Collaborations],
  Nucl.\ Phys.\ Proc.\ Suppl.\  {\bf 140} (2005) 832
  [hep-lat/0409003].

\bibitem{Asakawa:2000tr}
  M.~Asakawa, T.~Hatsuda and Y.~Nakahara,
  Prog.\ Part.\ Nucl.\ Phys.\  {\bf 46} (2001) 459
  [hep-lat/0011040].

\bibitem{Praki:2015yua}
  C.~Praki and G.~Aarts,
  PoS LATTICE {\bf 2015} (2015) 182
  [arXiv:1510.04069 [hep-lat]].


\bibitem{Albanese:1987ds}
  M.~Albanese {\it et al.}  [APE Collaboration],
  Phys.\ Lett.\ B {\bf 192} (1987) 163.






\end{thebibliography}
\end{document}